\documentclass[11pt,twoside]{article}

\usepackage{asp2006}
\usepackage{epsf}
\usepackage{lscape}

\begin{document}
\title{The Stellar Populations of dE galaxies in nearby Groups}   
\author{G. S. Da Costa}   
\affil{Research School of Astronomy \& Astrophysics, Australian National University, Mt Stromlo Observatory, Canberra, ACT, Australia}    

\begin{abstract} 
In this contribution initial results from colour-magnitude diagrams for five dEs in the M81 group
are presented.  The colour-magnitude diagrams, derived from HST/ACS images, reach well below 
the red giant branch tip and allow evaluation of distances and mean metallicities.  Further, the
intermediate-age upper-AGB stars seen in the diagrams allow estimates of the epochs of the
last significant episodes of the star formation in the dEs\@.  These epochs, and the relative numbers of upper-AGB to red giant branch stars, vary significantly from dE to dE\@.  Preliminary inferences from similar HST/ACS data for five early-type dwarfs in the Sculptor group are also discussed.
\end{abstract}


\section{Introduction}

One of the most important results in stellar populations research over the past decade or more
has been the recognition that, at least for systems within the Local Group, the objects classified as
dwarf Elliptical (dE) and dwarf Spheroidal (dSph) galaxies do not possess single old populations.
Instead they exhibit a variety of star formation histories.  This diversity
is best documented for the Milky Way Galaxy's (MWG) `classic' dSph companions, i.e.\ the well-studied 
objects as distinct from the numerous newly discovered, but as yet less well-studied, systems.  For
example, Ursa Minor is dominated by a single old stellar population but dSphs such as Carina, 
Fornax and Leo~I have had complex star formation histories, and contain stars as young as 
$\sim$1 Gyr.  Extended star formation histories are also evident for the M31 dE and dSph 
companions.  

At present there is no definitive understanding of what drives these variations in star 
formation histories, though there are hints that environment, as represented by, for example, proximity
to a luminous galaxy, and the type of that galaxy, plays a role.  These hints include the well known
morphology-density relation in which the majority (but not all!) of the isolated dwarf galaxies in
the Local Group are star-forming gas-rich dwarf irregulars, rather than dEs or dSphs, and that for
the MWG and its classic dSph companions, there is a trend for the more distant systems to have 
more prominent intermediate-age populations.  Clearly, to understand more fully the role of
environment, the next step is to investigate the stellar populations of dEs beyond the Local Group.
In this contribution initial results are presented for dEs in the M81 group, an environment 
somewhat denser than that of the Local Group, and for early-type dwarfs in the Sculptor Group, 
a low density loose
aggregation of galaxies strung out along the line-of-sight.  The results are based on images taken
with the ACS instrument aboard HST under programs GO-9884 and GO-10503.
\begin{figure}
\plotfiddle{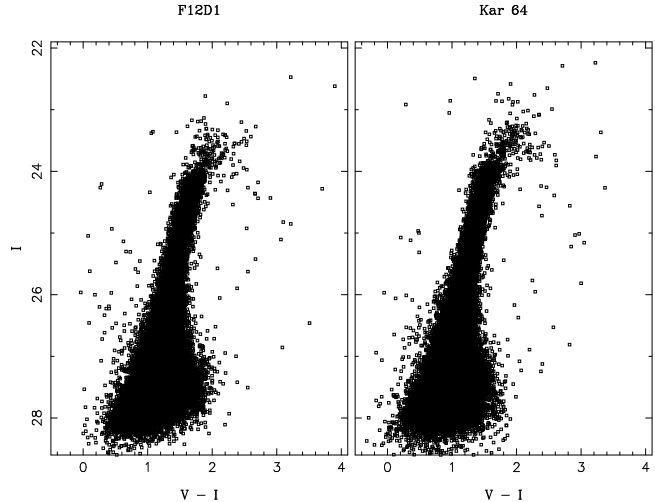}{6cm}{-90}{37}{37}{-150}{200}
\caption{Colour-magnitude diagrams for the M81 group dEs F12D1(left) and Kar~64 (right) based
on deep HST/ACS images.  The tip of the red giant branch is apparent at $I\approx24$ and there
are numerous upper-AGB stars at magnitudes above the RGB tip.  The CMDs for the other dEs
observed with HST/ACS are similar.  \label{m81_cmds}}
\end{figure}

\section{The M81 Group}

There are currently about a dozen known dE/dSph members of the M81 group \citep{Ka02}.   
Two, F8D1 and BK~5N, were studied in detail by \citet{NC98} using deep HST/WFPC2 images,
and deep HST/ACS images have now been obtained for a further five objects:
Kar~61, F6D1, Kar~64, F12D1 and DDO~71.  
These seven dEs cover a range of properties.  For example, the projected distances from M81
vary between $\sim$30kpc and $\sim$170kpc, and the absolute magnitudes range
from M$_{V}\approx-14.2$ to M$_{V}\approx-11.3$.  The HST/ACS images were reduced
with Stetson's DAOPHOT/ALLSTAR package \citep{PS87,PS94} and calibrated using the procedures outlined in \citet{Si05}.  \mbox{Fig.\ \ref{m81_cmds}} shows the resulting colour-magnitude diagrams 
(CMD) for F12D1 and Kar~64.  The CMDs for the other dEs are similar.  

In these CMDs the red giant branch (RGB) tip is readily apparent at
$I\approx24$, as is the presence of numerous upper-AGB stars at $I$ magnitudes above that
of the RGB tip.  Less obvious in these CMDs is a `red clump' of core helium burning stars
seen near the limit of the data at $I\approx27.5$ (M$_{I}\approx-0.3$).  Given the low metallicities of 
the dEs (see below), the presence of such a clump most likely indicates the existence of a population
of stars with ages younger than the Galactic globular clusters.  However, to recover information of
this nature, `population modelling' of the CMDs, in which the
observed data are compared to synthetic CMDs generated for different star formation histories,
needs to be undertaken.
Such comparisons require knowledge of completeness corrections and errors as a function of
magnitude and location, and this in turn requires substantial artificial star tests which have not yet
been completed.  Consequently, only results from the upper part of the CMDs where
the errors are small and the completeness corrections are negligible are presented.

\subsection{Distances}

\begin{figure}
\plotfiddle{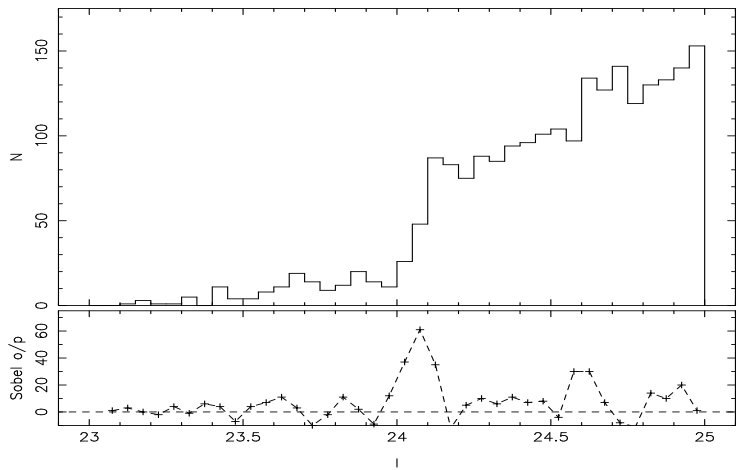}{6cm}{-90}{45}{35}{-180}{200}
\caption{Upper panel: $I$-band luminosity function ($I<25$ only) for the dE F12D1.  Lower panel:
output of a Sobel edge-detection filter applied to the data in the upper panel.  The value of 
$I(TRGB)$ is well determined by this process.  \label{ILF} }
\end{figure}

The distances were determined by constructing $I$-band luminosity functions (LF)
and measuring $I(TRGB)$, the $I$ magnitude of the tip of the red giant branch.  The process is
illustrated in Fig.\ \ref{ILF}, which shows the $I$-band LF for F12D1.  Application of a Sobel 
edge-detection filter to the LF data results in a well determined value for $I(TRGB)$\@.  Then
using reddenings from \citet{SFD98} and the calibration of $I(TRGB)$ from \citet{DA90} 
\citep[cf.][]{NC98} values for the distances then follow.   The mean distance of the 7
dEs is 3.70 Mpc, with a 1$\sigma$ dispersion of only 140 kpc, consistent with that expected 
just from the uncertainties in the $I(TRGB)$ values.  Indeed the difference in distance modulus
between the `nearest' and `furthest' galaxies is only $0.24\pm0.12$ mag, so that it is only at the 
$2\sigma$ level that all the dwarfs being at the same distance can be ruled out.   The mean
distance closely matches that of M81, 3.63 Mpc \citep{WF94}.  Clearly, the 7 dEs
lie within the `core' of the M81 group, consistent with the morphology - density relation.

\subsection{Abundances}

Once the distances to the dEs are known, estimates of the mean abundance can be derived
from the mean colour of the RGB compared to the colours of standard Galactic globular
cluster giant branches \citep[cf.][]{NC98}.  Strictly, abundances derived in this way are lower
limits on the actual mean abundance as, for fixed abundance, the RGB of a younger
population is bluer.  The effect is not large unless the bulk of the population is significantly
younger ($\Delta t > $ $\sim$6 Gyr) than the Galactic globular clusters, which is unlikely to be the 
case here.

\begin{figure}
\plotfiddle{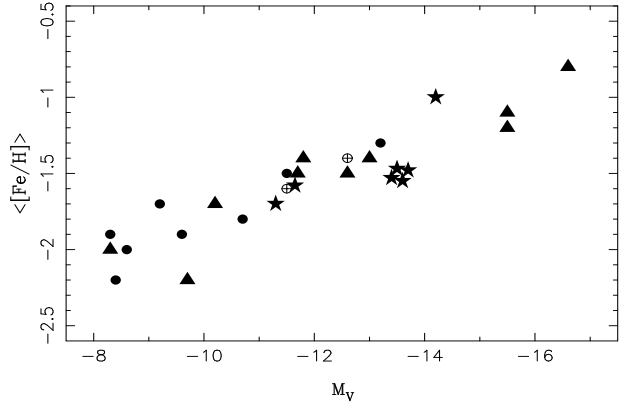}{5cm}{-90}{50}{45}{-180}{225}
\caption{Mean abundances plotted against absolute visual magnitudes for dEs in the M81
group (star symbols), in the Cen~A group (circled plus-signs) and the Local Group (Milky Way
Galaxy companions filled circles; M31 companions filled triangles). \label{LZplt} }
\end{figure}

In Fig.\ \ref{LZplt} the mean abundances derived in this way are plotted against the absolute
magnitudes of the systems.  Also shown are the data for the two 
Cen~A group dEs studied by \citet{Re06} as well as similar data for the classic
dSph companions to the Milky Way, and for dE and dSph companions to M31.  The M81 group
objects fall in with the Local Group systems, as do the Cen~A group objects, outlining
a well-defined relation.  The existence of this well-defined relation indicates that the 
mass-metallicity relation
for dE galaxies is not strongly dependent on environment, though clearly additional systems in
other environments need to be added to this relation before it can be claimed as `universal'.   The
relationship is usually explained in terms of supernova-driven gas loss within
dark matter halos \citep[e.g.][]{DW03}.

\subsection{Intermediate-Age Populations}

All seven M81 group dEs show populations of stars
above the RGB tip in their CMDs (cf.\ Fig.\ \ref{m81_cmds}).  Such stars are upper-AGB stars of 
intermediate-age (age $\sim1-10$ Gyr).  Two characteristics of 
these upper-AGB populations have been explored.  The first is the mean luminosity of the three most
luminous upper-AGB stars in each dE, $\langle M_{3}(bol) \rangle$.  Since the tip of the AGB is
more luminous for younger populations, this parameter is an indicator of the epoch of the
most recent episode of significant star formation.  Second, the number of upper-AGB stars has 
been compared to the number of RGB stars
in a 0.3 mag interval just below the RGB tip.  This parameter gives a first order indication of the 
relative importance of the intermediate-age population.

Based on the $\langle M_{3}(bol) \rangle$ values, and using the calibration of upper-AGB
tip luminosity with age of \citet{Re06}, the inferred epochs of last significant star formation for the
M81 group dEs vary from $\sim$2.5 to $\sim$8 Gyr.  The diversity in these values is similar to
that seen in the Local Group, though apparently as yet there are no M81 group dEs whose youngest
significant populations are as young as those seen in the MWG dSph companions Fornax
($<$1 Gyr) or Leo~I ($\sim$1 Gyr).   Although the most distant from M81 system (F8D1) is also
the dE with the most luminous upper-AGB stars, there is no clear dependance of the
$\langle M_{3}(bol) \rangle$ values on distance from M81, in contrast to the situation for the 
MWG classic dSph companions.   There the inner systems (R$_{GC}$ $\approx$ 75 kpc), such as 
Draco and Ursa Minor, lack upper-AGB stars and so have $\langle M_{3}(bol) \rangle$ values
of order --3.6, the RGB tip luminosity.  On the other hand, the more distant systems such as
Fornax at $\sim$140 kpc and Leo~I at $\sim$250 kpc have $\langle M_{3}(bol) \rangle$
values of --5.0 and --4.9, respectively, so that overall there is an apparent correlation between 
$\langle M_{3}(bol) \rangle$ and R$_{GC}$ for the MWG classic dSph companions.  This correlation 
is a manifestation of the suggestion originally by \citet{vdB9x}
that the Milky Way galaxy has more strongly influenced its nearer neighbours compared to the more
distant satellites.  A comparable trend is much less obvious for the M31 dSph companions, though the
only M31 dSph companion with an upper-AGB population is amongst the most distant of the
satellites \citep[cf.][]{DA00}.  The situation for M81 is thus more similar to that for M31 than for the
MWG.

As regards the relative numbers of upper-AGB stars compared to the number of RGB stars just
below the tip, there are definite dE-to-dE variations.  Again
there is no clear correlation of the upper-AGB to RGB number ratio with distance from M81, though 
the most distant from M81 dE is once more the one with the highest value of this ratio.  Full population
synthesis models are needed to infer the proper intermediate-age population fractions but
application of simple models \citep[cf.][]{Re06}  suggests that these fractions range from $\sim$15\%
to perhaps $\sim$50\%.  Once again these values fall within those for MWG dSph companions 
which range from $\sim$0\% (e.g.\ Draco, Ursa Minor) to $\sim$70--80\% \citep[Leo~I;][]{Ga99}.

\section{The Sculptor Group}

The Sculptor group contains six early-type dwarf galaxies, once of which is the relatively luminous
(M$_{V}\approx-15.3$) dS0 galaxy NGC~59 that will not be discussed here.  Regarding the other
five systems (ESO540-030, ESO540-032, ESO410-005, ESO294-010 and Scl-dE1) recent 
observations have detected neutral gas in four, with M$_{HI}$/L$_{B}$ values 
between 0.08 and
0.18 in solar units \citep{AB05}.  These systems may well be best regarded as `transition objects' 
rather than as dE systems.  The fifth galaxy, Scl-dE1 has M(H{\small I}) $<$ 10$^{5}$ solar masses and 
M$_{HI}$/L$_{B}$ $<$ 0.04 in solar units \citep{AB05}.

In the past few months deep HST/ACS observations of the five dwarfs have been obtained.  
The images will
allow generation of CMDs of similar quality to those for the M81 group dEs presented in the previous
section.  A first look at the data confirms that the five Scl group dwarfs have a significant
spread in distance: Scl-dE1 is the most distant, ESO540-030 and ESO540-032 are at similar intermediate distances, and ESO294-010 and ESO410-005 are the closest.   All 5 systems appear to contain upper-AGB stars, and there are apparent differences in the upper-AGB populations from system to system.  While the CMD data are far from fully analyzed, it is likely that the red clump/horizontal 
branch is reached in the two nearer systems.  The intention is to search the datasets for RR Lyrae variables to confirm the existence of a Galactic globular cluster aged population in the two dwarfs.

The above results must be considered preliminary as the initial CMDs are not yet calibrated, nor has
there been any attempt to correct them for field contamination or to remove spurious or poorly 
measured stars.  One issue though can be
investigated directly from the ACS images.  This the question of the existence of globular clusters 
associated with Scl-dE1.
\citet*{Sh05} identified three candidate globular clusters based on relatively
short exposure HST/WFPC2 `snapshot' images.   
All three candidates are contained within the field of the ACS images, which have better resolution 
and which go considerably deeper than the images available to \citet{Sh05}.  Inspection of the  
images indicates that all three of the \citet{Sh05} candidates are in fact background 
galaxies.  Nevertheless, the ACS images also clearly show that Scl-dE1 does possess at least
one apparently 
genuine globular cluster candidate.  The cluster candidate lies $\sim$20$^{\prime\prime}$ or 
420pc to the NW of the galaxy centre and is $\sim$50pc in diameter.  
It should be possible to produce a CMD for the cluster candidate which
can then be compared with that for the dwarf.  The ACS frames of the other dwarfs will also be surveyed
for cluster candidates.

\section{Conclusions}

While fully exploiting all the information contained in the deep CMDs of dEs in the Scl and M81
groups remains a task to be completed, it is clear that essentially all of the systems studied so
far contain indications of extended star formation histories.  Stated in another way, it is becoming 
increasingly
evident that dEs with stellar populations similar to that of the Local Group dEs Ursa Minor and 
Tucana, which lack upper-AGB stars and which have dominant old (age $>$ 10 Gyr) populations,
are not very common.   ``Classic'' dEs, meaning objects consisting of a single 
ancient stellar population (cf.\ Baade Pop II) are apparently comparatively rare objects.


\acknowledgements  I would like to thank all my collaborators, particularly Nelson Caldwell, for 
their contributions to the material presented here.  This research has been supported in part by the 
Australian Research Council through grant DP0343156.



\begin{thebibliography}{}

\bibitem[Bouchard et al.(2005)]{AB05} Bouchard, A., Jerjen, H., Da~Costa, G. S., \& Ott, J. 2005, AJ, 130, 2058
\bibitem[Caldwell et al.(1998)]{NC98} Caldwell, N., Armandroff, T. E., Da~Costa, G. S., \& Seitzer, P.
1998, AJ, 115, 535
\bibitem[Da~Costa \& Armandroff(1990)]{DA90} Da~Costa, G. S., \& Armandroff, T. E. 1990, AJ, 100, 162
\bibitem[Da~Costa et al.(2000)]{DA00} Da~Costa, G. S., Armandroff, T. E., Caldwell, N., \& Seitzer, P.
2000, AJ, 119, 705
\bibitem[Dekel \& Woo(2003)]{DW03} Dekel, A., \& Woo, J. 2003, MNRAS, 344, 1131
\bibitem[Freedman et al.(1994)]{WF94} Freedman, W. L., et al. 1994, ApJ, 427, 628
\bibitem[Gallart et al.(1999)]{Ga99} Gallart, C., Freedman, W. L., Aparicio, A., Bertelli, G., \& Chiosi, C.
1999, AJ, 118, 2245
\bibitem[Karachenstev et al.(2002)]{Ka02} Karachentsev, I., et al. 2002, A\&A, 383, 125
\bibitem[Rejkuba et al.(2006)]{Re06} Rejkuba, M., Da Costa, G. S., Jerjen, H., Zoccali, M., \&
Binggeli, B. 2006, A\&A, 448, 983
\bibitem[Schlegel et al.(1998)]{SFD98} Schlegel, D. J., Finkbeiner, D. P., \& Davis, M. 1998, ApJ, 500,
525
\bibitem[Sharina et al.(2005)]{Sh05} Sharina, M. E., Puzia, T. H., \& Makarov, D. I. 2005,
A\&A, 442, 95
\bibitem[Sirianni et al.(2005)]{Si05} Sirianni, M., et al. 2005, PASP, 117, 1049
\bibitem[Stetson(1987)]{PS87}Stetson, P. B. 1987, PASP, 99, 191
\bibitem[Stetson(1994)]{PS94}Stetson, P. B. 1994, PASP, 106, 250
\bibitem[van den Bergh(1994)]{vdB9x} van den Berg, S. 1994, ApJ, 428, 617

\end{thebibliography}
\end{document}